\documentclass[twocolumn,showpacs,prl]{revtex4}
\usepackage{graphics}
\usepackage{color}

\begin{document}

\title{A maximum entropy principle explains
  quasi-stationary states in systems with long-range interactions:
 the example of the Hamiltonian Mean Field model}

\author{Andrea Antoniazzi$^{1}$\thanks{andrea.antoniazzi@unifi.it},
Duccio Fanelli$^{1,2}$\thanks{duccio.fanelli@ki.se}, Julien
Barr{\'e}$^{3}$\thanks{jbarre@math.unice.fr},
Pierre-Henri Chavanis$^{4}$\thanks{chavanis@irsamc.ups-tlse.fr}, Thierry
Dauxois$^{5}$\thanks{Thierry.Dauxois@ens-lyon.fr}, Stefano
Ruffo$^{1}$\thanks{stefano.ruffo@unifi.it}}

\affiliation{ 1. Dipartimento di Energetica and CSDC, Universit\`a di
  Firenze, and INFN, via S. Marta, 3, 50139 Firenze, Italy\\
  2. Department of Cell and Molecular Biology, Karolinska Institute,
  SE-171 77 Stockholm, Sweden \\ 3. Laboratoire J. A. Dieudonn\'e, UMR
  CNRS 6621, Universit\'e de Nice Sophia-Antipolis Parc Valrose 06108
  Nice c\'edex 2, France\\ 4. Laboratoire de Physique
  Th\'eorique, Universit\'e Paul Sabatier, 118,
  route de Narbonne 31062 Toulouse, France\\ 5. Laboratoire de
  Physique, UMR CNRS 5672, ENS Lyon, 46 All\'{e}e d'Italie, 69364 Lyon
  c\'edex 07, France} \date{\today}

\begin{abstract}
A generic feature of systems with long-range interactions is the
presence of {\it quasi-stationary} states with non-Gaussian single
particle velocity distributions.  For the case of the Hamiltonian Mean
Field (HMF) model, we demonstrate that a maximum entropy principle
applied to the associated Vlasov equation explains known features of
such states for a wide range of initial conditions.
We are able to reproduce velocity distribution functions with an
analytical expression which is derived from the theory with no
adjustable parameters. A normal diffusion of angles is detected and a new
dynamical effect, two oscillating clusters surrounded by a halo, is
also found and theoretically justified. 
\end{abstract}


\maketitle

Long-range interactions are common in nature~\cite{Houches02}.
Examples include: self gravitating systems~\cite{Padmanabhan},
plasmas~\cite{Nicholson}, dipolar magnets \cite{Akhiezer} and
wave-particle interactions~\cite{Barre}.  Theoretical studies have
shown that the thermodynamic properties of these systems differ from
those of systems with short-range interactions.  For instance,
long-range interactions may produce a negative microcanonical specific
heat~\cite{LyndenBell68} and, more generally, inequivalence of the
canonical and microcanonical ensembles~\cite{Mukamel}.  Also the
dynamics of models with long-range interactions has been studied,
revealing a variety of peculiar features such as the presence of
breaking of ergodicity in microcanonical
dynamics~\cite{Celardo,Schreiber} and the existence of
quasi-stationary states whose relaxation time to equilibrium diverges
with system size~\cite{antoni-95,Taruya}.  A number of paradigmatic
toy models have been proposed that provide the ideal ground for
theoretical investigations. Among others, the Hamiltonian Mean Field
(HMF) model~\cite{antoni-95} is nowadays widely analyzed because it
displays many features of long range interactions, while being
simple to study analytically and numerically.  
Within the
HMF scenario, non-Gaussian velocity distributions~\cite{Rapisarda} and
signatures of anomalous diffusion~\cite{Latora} have been reported in
the literature. These discoveries have originated an intense debate
about the general validity of Boltzmann-Gibbs statistical mechanics
for systems with long-range interactions~\cite{EPN}.  Non-Gaussian
distributions have been {\em fitted} using Tsallis' $q$--exponentials
\cite{Tsallis},
i.e. algebraically decaying profiles predicted within the realm of
nonextensive statistical mechanics. Based on these findings, the
generalized thermostatistic formulation pioneered by Tsallis
was proposed as a tool to describe
the properties of quasi-stationary states\cite{Rapisarda}.

In this Letter, we demonstrate that a maximum entropy principle
inspired by Lynden-Bell's theory of ``violent relaxation" for the
Vlasov equation allows to explain satisfactorily the
numerical simulations performed for the HMF model. Analytically
obtained PDF's are superimposed to the numerics {\em without}
adjusting any free parameter.  In other words, our results point to
the fact that there is no need to invoke generalized forms of
Boltzmann-Gibbs statistical mechanics to describe the non-equilibrium
properties of this system.

The HMF model describes the motion of $N$ coupled rotators and is
characterized by the following Hamiltonian
\begin{equation}
\label{eq:ham}
H = \frac{1}{2} \sum_{j=1}^N p_j^2 + \frac{1}{2 N} \sum_{i,j=1}^N 
\left[1 -  \cos(\theta_j-\theta_i) \right]
\end{equation}
where $\theta_j$ represents the orientation of the $j$-th rotor and
$p_j$ is its conjugate momentum. To monitor the evolution of the
system, it is customary to introduce the magnetization, a global order
parameter defined as $M=|{\mathbf M}|=|\sum {\mathbf m_i}| /N$, where
${\mathbf m_i}=(\cos \theta_i,\sin \theta_i)$ is the local
magnetization vector. Starting from {\it out--of--equilibrium} initial
conditions, the system gets trapped in Quasi-Stationary States (QSS),
whose lifetime diverges when increasing the number of particles
$N$. Importantly, when performing the mean-field limit ($N
\rightarrow \infty$) {\it before} the infinite time limit, the system
cannot relax towards Boltzmann--Gibbs equilibrium and remains
permanently confined in QSS. In this regime, the magnetization is
lower than the one predicted by the Boltzmann--Gibbs equilibrium and the
system apparently displays a number of intriguing anomalies, e.g. non
Gaussian velocity distributions~\cite{Rapisarda} and non standard
diffusion in angle~\cite{Latora}. We shall here provide a strong
evidence that the above phenomena can be successfully interpreted in
the framework of the statistical theory of the Vlasov equation, a
general approach originally introduced in the astrophysical and 2D
Euler turbulence contexts~\cite{LyndenBell67,Chavanis96,PhysAJulien}.

First, let us recall that for mean-field Hamiltonians such
as~(\ref{eq:ham}), it has been rigorously
proven~\cite{BraunHepp} that, in the $N \to \infty$ limit, the
$N$-particle dynamics is described by the Vlasov equation
\begin{equation}
\frac{\partial f}{\partial t} + p\frac{\partial f}{\partial \theta} -
\frac{d V}{d \theta} \frac{\partial f}{\partial p}=0\quad ,
\label{eq:VlasovHMF}
\end{equation}
where $f(\theta,p,t)$ is the microscopic one-particle
distribution function and
\begin{eqnarray}
V(\theta)[f] &=& 1 - M_x[f] \cos(\theta) - M_y[f] \sin(\theta) ~, \\
M_x[f] &=& \int_{-\pi}^{\pi} \int_{-\infty}^{\infty}  f(\theta,p,t) \, \cos{\theta}  {\mathrm d}\theta
{\mathrm d}p\quad , \\
M_y[f] &=& \int_{-\pi}^{\pi} \int_{\infty}^{\infty}  f(\theta,p,t) \, \sin{\theta}{\mathrm d}\theta
{\mathrm d}p\quad .
\label{eq:pot_magn}
\end{eqnarray}
The specific energy $h[f]=\int \int (p^2/{2}) f(\theta,p,t) {\mathrm d}\theta
{\mathrm d}p - ({M_x^2+M_y^2 - 1})/{2}$ and momentum
$P[f]=\int \int p f(\theta,p,t) {\mathrm d}\theta
{\mathrm d}p$ functionals are conserved quantities.



We now turn to illustrate the maximum entropy method.  The basic idea
is to coarse-grain the microscopic one-particle distribution function
$f(\theta,p,t)$ into a given set of values. It is then possible to
associate an entropy to the coarse-grained distribution $\bar{f}$,
i.e. averaged over a box of finite size, and statistical equilibrium
can be determined by maximizing this entropy while imposing the
conservation of certain Vlasov dynamical invariants.
A physical description of this procedure can be found
in \cite{LyndenBell67,Chavanis06} and a rigorous mathematical
justification is given in Ref.~\cite{Michel94}.

In the following, we shall assume that the initial single particle
distribution takes only two distinct values, namely $f_0=1/(4
\Delta_{\theta} \Delta_{p})$, if the angles (velocities) lie within an
interval centered around zero and of half-width $\Delta_{\theta}$
($\Delta_{p}$), and is zero otherwise. This choice corresponds to the
so-called ``water-bag" distribution which is fully specified by energy
$h[f]=e$, momentum $P[f]=\sigma$ and initial magnetization ${\mathbf
M_0}=(M_{x0}, M_{y0})$.  Vlasov time evolution can modify the shape of
the boundary of the ``water-bag", while conserving the area inside
it. Hence, the distribution remains two-level ($0,f_0$) as time
progresses. Coarse-graining amounts to performing a local average of $f$
inside a given box and this procedure results in $\bar{f}$. In this two-level
situation, the
mixing entropy per particle associated with $\bar{f}$ reads
\begin{equation}
\label{entropy_shape}
s(\bar{f})=-\int \!\!{\mathrm d}p{\mathrm d}\theta \,
\left[\frac{\bar{f}}{f_0} \ln \frac{\bar{f}}{f_0}
+\left(1-\frac{\bar{f}}{f_0}\right)\ln
\left(1-\frac{\bar{f}}{f_0}\right)\right].
\end{equation}
The shape of this entropy derives from a simple combinatorial 
analysis\cite{LyndenBell67,Chavanis06}.
The maximum entropy
principle is then defined by the following constrained variational
problem
\begin{eqnarray}
S(e,\sigma)\! =\! \max_{\bar{f}} \biggl(\!  s(\bar{f})
\biggr|
 h(\bar{f})=e;\;\!\! P(\bar{f})=\sigma;\; \!\!\!
\int \!\! {\mathrm d}\theta {\mathrm d}p \bar{f}=1\biggr)~.
\label{eq:problemevar}
\end{eqnarray}
The problem is solved by introducing three Lagrange multipliers
$\beta/f_0$, $\lambda/f_0$ and $\mu/f_0$ for energy, momentum and
normalization. This leads to the following analytical form of the
distribution
\begin{equation}
\label{eq:barf} \bar{f}(\theta,p)= f_0\frac{e^{-\beta (p^2/2
- M_y[\bar{f}]\sin\theta
- M_x[\bar{f}]\cos\theta)-\lambda p-\mu}}
{1+e^{-\beta (p^2/2  - M_y[\bar{f}]\sin\theta
 - M_x[\bar{f}]\cos\theta)-\lambda p-\mu}}.
\end{equation}

\begin{figure}[t]
\resizebox{0.5\textwidth}{!}{\includegraphics{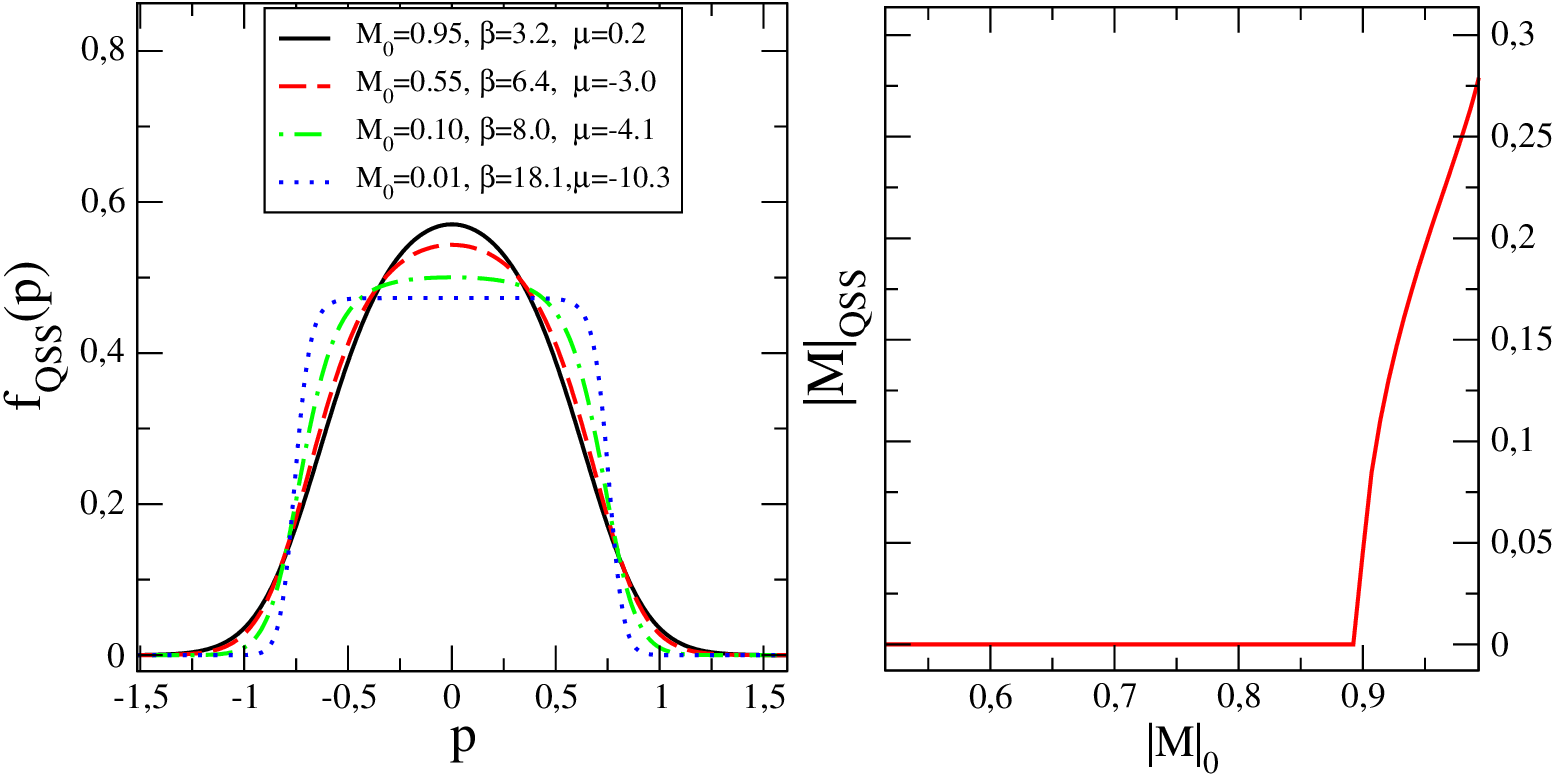}}
\caption{ \label{fig1} a) Velocity profiles~(\ref{eq:veloc_distr})
predicted theoretically for different initial magnetizations (see
legend, where we have reported also the values of the parameter $\beta$
 and $\mu$). b) final magnetization as a function of
$M_0$. A phase transition is observed at $M_0=0.897$ for
$e=0.69$. }
\end{figure}

This distribution differs from the Boltzmann-Gibbs one because of the
``fermionic'' denominator which is originated by the form
(\ref{entropy_shape}) of the entropy.
Inserting expression (\ref{eq:barf}) into the energy, momentum and
normalization constraints and using the definition of
magnetization, it can be straightforwardly shown that the momentum
multiplier $\lambda$ vanishes. Moreover, defining $x=e^{-\mu}$ and
${\mathbf m}=(\cos \theta, \sin \theta)$, yields the following system
of implicit equations in the unknowns $\beta$, $x$, $M_x$ and $M_y$
\begin{eqnarray}
\label{eq:cond0}
&f_0& \frac{x}{\sqrt{\beta}} \int {\mathrm d} \theta e^{\beta {\mathbf
M} \cdot {\mathbf m}} F_0\left(x e^{\beta {\mathbf M} \cdot {\mathbf
m}}\right) = 1 \\
\label{eq:cond3}
&f_0& \frac{x}{2 \beta^{3/2}} \int {\mathrm d} \theta
e^{\beta {\mathbf M} \cdot {\mathbf m}}
F_2\left(x e^{\beta {\mathbf M} \cdot {\mathbf m}}\right)
= e+\frac{M^2 - 1}{2} \nonumber\\
\label{eq:cond1}
&f_0& \frac{x}{\sqrt{\beta}} \int {\mathrm d} \theta \cos \theta
e^{\beta {\mathbf M} \cdot {\mathbf m}}
F_0\left(x e^{\beta {\mathbf M} \cdot {\mathbf m}}\right)
= M_x \nonumber \\
\label{eq:cond2}
&f_0& \frac{x}{\sqrt{\beta}} \int {\mathrm d} \theta \sin \theta
e^{\beta {\mathbf M} \cdot {\mathbf m}}
F_0\left(x e^{\beta {\mathbf M} \cdot {\mathbf m}}\right)
= M_y \nonumber
\end{eqnarray}
with $F_0(y) = \int \exp (-v^2/2)/(1+y \exp (-v^2/2)){\mathrm d}v$,
$F_2(y) = \int v^2 \exp (-v^2/2)/(1+y \exp (-v^2/2)){\mathrm d}v$,
where $v=\sqrt{\beta}p$.
This system of equations is then solved using a Newton-Raphson method
and the integrals involved are also performed numerically.
For $e=\lim_{N\rightarrow \infty}H/N=0.69$, a value often considered in the
literature \cite{rapisarda}, the maximum entropy state has zero
magnetization, for {\em any initial magnetization} $M_0=|{\mathbf
M_0}|<M_{crit}=0.897$. Hence, the QSS distribution does not
depend on the angles and the velocity distribution
can be simplified into
\begin{equation}
\label{eq:veloc_distr}
f_{QSS}(p)=
f_{0}\frac{e^{-\beta p^2/2-\mu}}
{1+e^{-\beta p^2/2-\mu}},
\end{equation}
with $\beta$ and $\mu$ numerically determined from system
(\ref{eq:cond0}). Velocity profiles predicted by
(\ref{eq:veloc_distr}) are displayed in Fig.~\ref{fig1}a for different
values of the initial magnetization. Gaussian tails are always
present, contrary to the power-law ($q$-exponential) fits reported in
Ref.~\cite{Rapisarda}.  Note that the power-law decay was already
excluded on the basis of numerical simulations in Ref.~\cite{yama} for
initial zero magnetization states.  At
$M_0=M_{crit}=0.897$, a bifurcation occurs (see Fig.~\ref{fig1}b) and
the magnetization of the quasi-stationary state $M_{QSS}$ becomes non
zero, which means that the equilibrium Lynden-Bell distribution
develops an inhomogeneity in angles. The details of
this phase transition are further discussed in Ref. \cite{pthmf}.

\begin{figure}[th]
\vskip 0.5truecm
\resizebox{0.4\textwidth}{!}{\includegraphics{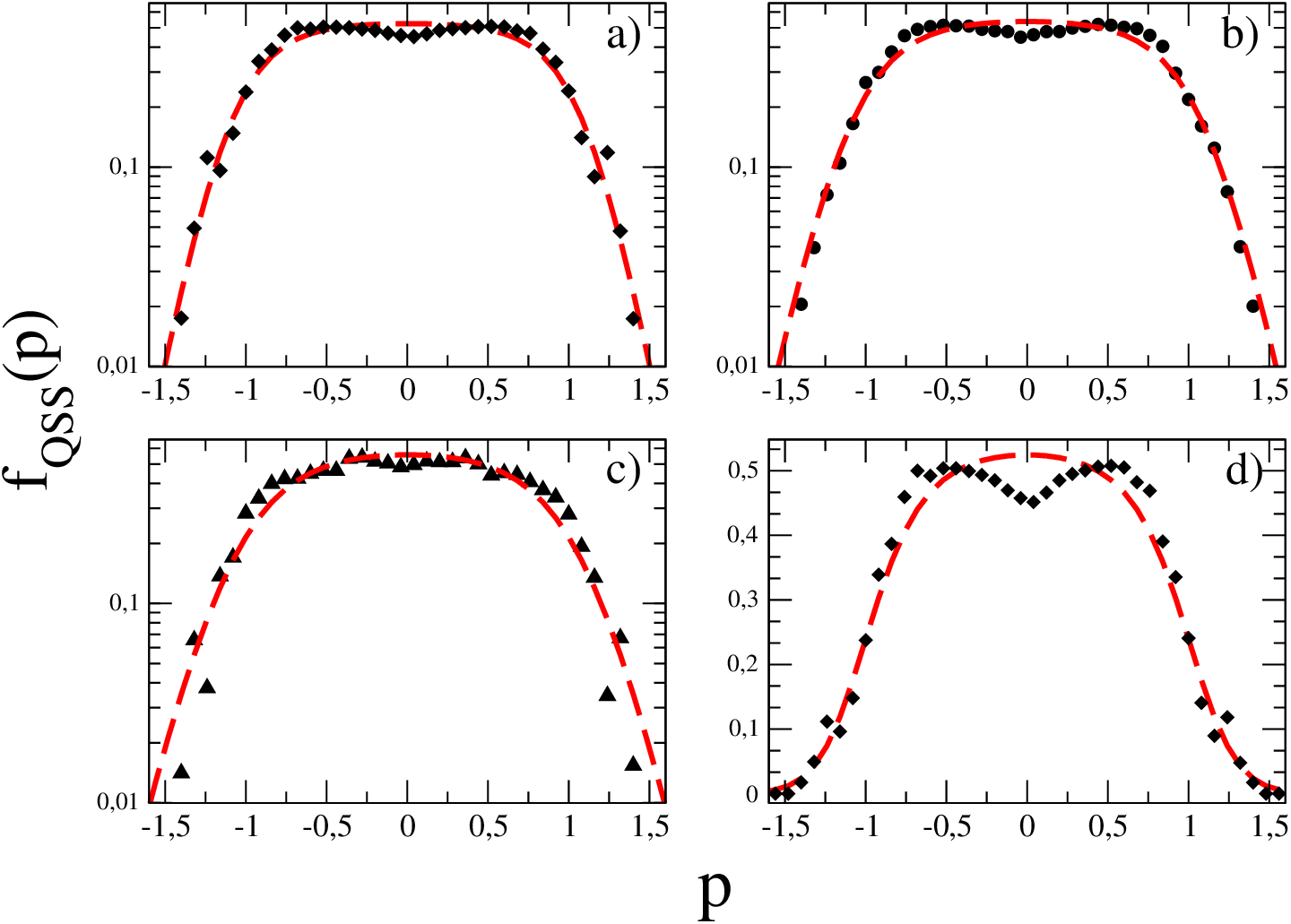}}
\caption{\label{fig2} Velocity distribution functions. Symbols refer
to numerical simulations, while dashed lines stand for the
theoretical profile (\ref{eq:veloc_distr}). Panels a), b) and c)
present the three cases ${M_0}=0.3$, ${M_0}=0.5$ and ${M_0}=0.7$ in
lin-log scale, while panel d) shows the case ${M_0}=0.3$ in lin-lin
scale.  The numerical curves are computed from one single realization
with $N=10^7$ at time $t=100$. Here $e=0.69$.  }
\end{figure}

We validate our theoretical findings in the initial
magnetization range $M_0 \in [0,M_{crit}]$, by performing numerical
simulations with $N$ ranging from $10^3$ to $10^7$. Numerical velocity
distributions are compared in Fig.~\ref{fig2} with the analytical
solution~(\ref{eq:veloc_distr}).  Although not a single free parameter
is used, we find an excellent agreement in the tails of the
distribution. The discrepancies observed in the center of the
distributions are commented below.

To discuss the behavior of $M$ in the QSS, one must distinguish
different magnetization intervals. Consider first the interval $M_0
\in [M_a,M_{crit}]$, with $M_a \approx 0.5$.  Both $M_x$ and $M_y$ are
found to approach zero when the number of rotators is increased, in
agreement with the theory outlined above.  Our results correlate well
with the scaling $|M| \propto N^{-1/6}$ reported in
Ref.~\cite{rapisarda}.  Numerical simulations also confirm the
presence of a bifurcation at $M_0 = M_{crit}$, and indicate that the
distribution in angles is indeed inhomogeneous above this value.
Interestingly, when the initial magnetization lies instead in the
interval $[0,M_a]$, $M_x$ and $M_y$ display regular oscillations in
time, which appear only when a large enough number of rotators ($N >
10^6$) is simulated. It is important to emphasize that the
oscillations are centered around zero, i.e. the equilibrium value
predicted by our theory.  



We now discuss the presence of two symmetric bumps in
the velocity distributions obtained numerically (see
Fig.~\ref{fig2}).  This is a consequence of a collective phenomenon
which leads to the formation of two clusters in the $(\theta,p)$
plane.  Both clusters form early in time and then acquire constant
opposite velocities which are maintained during the time
evolution, thus enlarging their relative separation. Consequently, the
bumps displayed by the velocity distributions are not transient
features, but represent instead an intrinsic peculiarity of QSS.  
To the best of our knowledge, this is a new collective 
phenomenon which has not been
previously detected.  A simple dynamical argument can be elaborated to
shed light onto the process of formation of the clusters. Consider the
one-particle Hamiltonian $\epsilon(\theta, p) = \frac{p^2}{2} - M_x
\cos \theta - M_y \sin \theta$ associated to (\ref{eq:ham}), where
$(\theta,p)$ are the conjugate variables of the selected rotor. For
short times, $\theta \sim \theta_{0} + p_{0} t$.  One then finds $M_x
\simeq (\sin{\Delta_{\theta}} \sin{\Delta_p t})/(\Delta_{\theta} \Delta_{p} t)$ and $M_y
\simeq 0$. Using this result, one ends up with
\begin{equation}
\label{ham_sps}
\epsilon(\theta, p) = \frac{p^2}{2} +\frac{\sin{\Delta_{\theta}}}{2
  \Delta_{\theta} \Delta_{p} t } \left[  \sin \left( \theta-
  \Delta_{p}t \right) +   \sin \left( \theta+
  \Delta_{p}t \right)  \right]~,
\nonumber
\end{equation}
which corresponds to the Hamiltonian of one particle interacting with
two waves of phase velocities $\pm \Delta_{p}$. Depending on the
initial condition, the particles can be trapped in one of the two
resonances, the latter being therefore directly responsible for the
arising of two highly populated regions. 
Moreover, by including higher order corrections to the above calculation
one can show that the two resonances tend to overlap 
when $M_0 \rightarrow 1$, in agreement with
our numerical findings.

Having derived an analytical expression for the velocity distribution
function~(\ref{eq:veloc_distr}), which is fully validated by the
numerics, enables to take advantage of the predictions recently
obtained in Ref.~\cite{BouchetDauxois}, where it has been demonstrated that 
momentum autocorrelation functions can be deduced by knowing only the
tails of the velocity distributions.  Since these are Gaussian 
(see Eq.~(\ref{eq:veloc_distr})), one expects algebraic decay of momentum
autocorrelation.
 The mean square displacement of the angles $\sigma^2(t)=\frac{1}{N}\sum_i{[\theta_i(t) -
\theta_i(0)]^2}$ is also a quantity of interest. The scaling
$\sigma^2 \propto t^{\gamma}$ defines the diffusive behavior:
$\gamma=1$ corresponds to normal diffusion and $\gamma=2$ to free
particle ballistic dynamics.  Intermediate cases correspond to the
anomalous diffusion behavior. Here, for all water-bag initial
conditions, the behavior of the momentum autocorrelation function
implies~\cite{BouchetDauxois} that a weakly anomalous diffusion has to
be expected, with a diffusion exponent $\gamma=1$ and logarithmic
corrections.
\begin{figure}[ht]
\vskip 0.2truecm
\resizebox{0.3\textwidth}{!}{\includegraphics{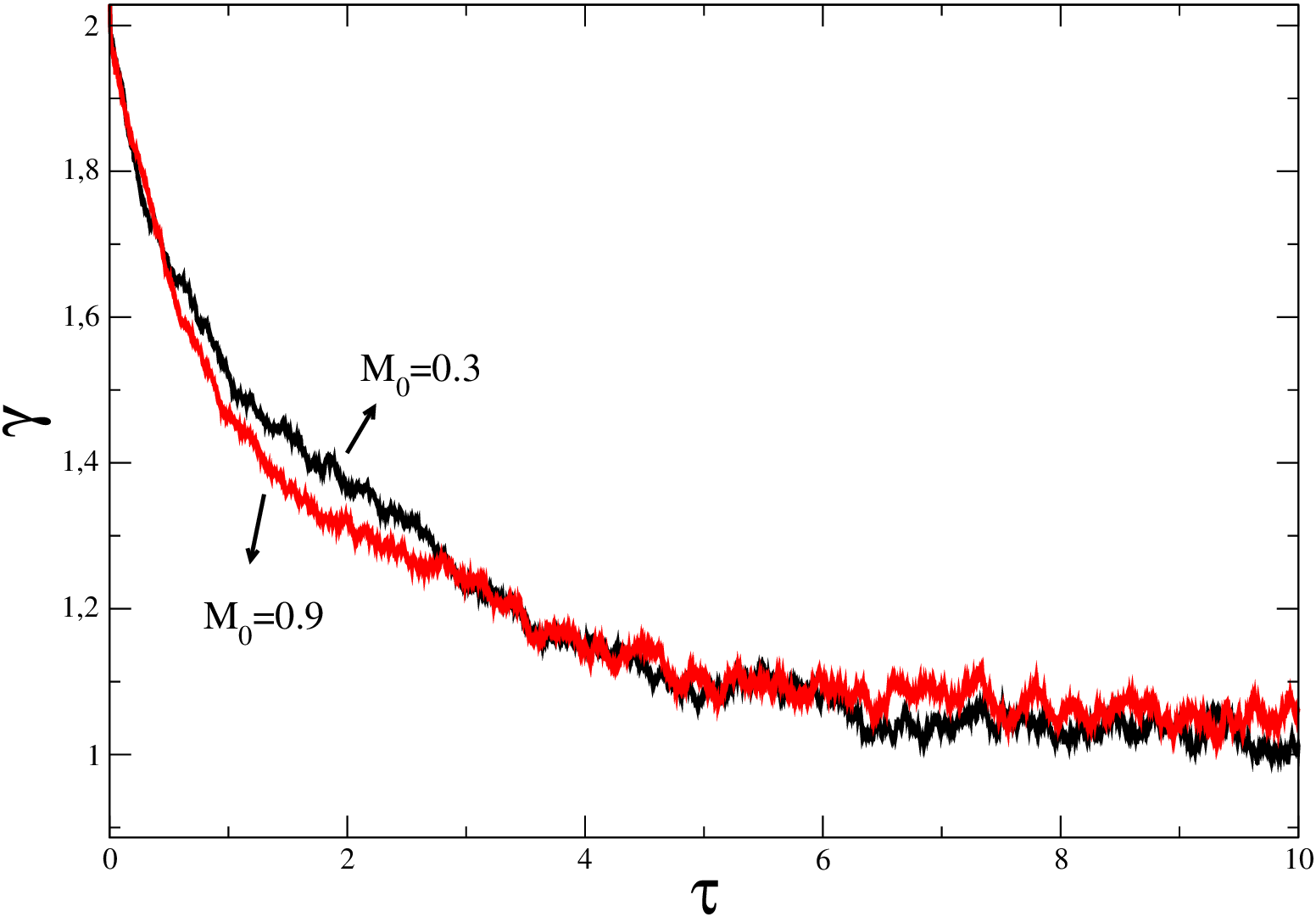}}
\caption{The exponent $\gamma={\rm d} \log(\sigma^2) / {\rm d}\log
(t)$ is plotted as a function of the rescaled time $\tau=t/N$.
Starting from the initial ballistic value $2$, it converges to the
normal diffusion exponent $1$. Simulations refer to $M_0=0.3$, and
$M_0=0.9$. Here $N=10^5$ and $e=0.69$.  }\label{fig3}
\end{figure}
On the numerical simulations side, it is claimed in
Ref.~\cite{rapisarda} that the QSS display anomalous diffusion
with an exponent $\gamma$ in the range $1.4-1.5$ for $0.4 \le M_0 \le
1$.  These results are contradicted by more recent papers
\cite{yama1}, where normal diffusion behavior is found. To provide
further insight, we have monitored the time evolution of $\sigma^2$ by
employing a larger number of particles than in previous
investigations.  As clearly demonstrated by the inspection of
Fig.~\ref{fig3}, a large value of the exponent $\gamma$ is clearly
excluded. On the contrary, the almost normal diffusion found is in
complete agreement with the theoretical scenario discussed above.

In this paper, by drawing analogies with the statistical theory of
``violent relaxation" in astrophysics and 2D Euler turbulence, we have
analytically derived known properties of the quasi-stationary states
of the HMF model. In particular: $i)$ velocity probability
distributions in all quasi-stationary states
investigated are well described by Lynden-Bell statistics
(\ref{eq:veloc_distr}); $ii)$ Gaussian tails of such maximum entropy
states ensure an algebraic decay of momentum autocorrelation functions
and, hence, a normal diffusion of the angles.  Our theoretical
approach is {\it fully predictive}, contrary to results obtained using
nonextensive thermostatistics\cite{Tsallis}, which consist in {\it parametric fits}
that are not justified from first principles\cite{Rapisarda}.  Despite
this success of ``violent relaxation'' theory, we do not expect it to
be so precise in all long-range systems: due to incomplete relaxation
of the Vlasov equation
\cite{LyndenBell67,Chavanis96,Chavanis06}, the QSS should
deviate somewhat from Lynden-Bell's statistical
prediction. Besides that, we have discovered that a double cluster
spontaneously forms for all magnetizations.  This collective effect
was not known before. Similar time-dependent quasi-stationary states
have been recently found in Ref~\cite{Kaneko}. Our maximum entropy principle is unable to
capture this phenomenon, for which we have developed an analytical
approach based on analogies with similar effects encountered in
plasma-wave Hamiltonian dynamics. More refined maximum entropy
schemes, accounting for the conservation of additional invariants
besides the normalization, are expected to give a full description of these
phenomena and represent a challenge for future investigations.

We acknowledge financial support from the PRIN05-MIUR project {\it
Dynamics and thermodynamics of systems with long-range interactions}.

\end{document}